\documentclass[prd,onecolumn,nofootinbib,showkeys]{revtex4}
\usepackage{bm,amsmath,amssymb,graphicx}

\begin{document}

\title{PHANTOM DARK ENERGY FROM SCALAR--TORSION COUPLING}
\author{Nikodem J. Pop{\l}awski}
\affiliation{Department of Mathematics and Physics, University of New Haven, 300 Boston Post Road, West Haven, CT 06516, USA}
\email{NPoplawski@newhaven.edu}

\noindent
{\em Modern Physics Letters A}\\
Vol. {\bf 35}, No. 40, 2050331 (2020)
\vspace{0.4in}

\begin{abstract}
We show that a scalar field without a kinetic term in the Lagrangian density, coupled to the covariant divergence of the torsion vector in the Einstein--Cartan theory of gravity, becomes kinetic in its general-relativistic equivalent formulation.
The resulting kinetic term is negative: such a scalar field could be a source of phantom dark energy.
\end{abstract}
\maketitle

The Einstein--Cartan theory of gravity \cite{KS,rev} naturally extends general relativity by including the spin angular momentum of matter.
The spin--spin interaction arising in this theory \cite{HD,Niko} may also remove divergent integrals in quantum field theory by providing fermions with spatial extension \cite{non}, avoid the formation of singularities from fermionic matter in black holes and in cosmology \cite{cosm}, and explain the origin of cosmic inflation \cite{ApJ}, the matter--antimatter asymmetry \cite{anti}, a cosmological constant \cite{affine}, and primordial fluctuations \cite{fluct}.
In this theory, the metric--affine Lagrangian density for the gravitational field is proportional to the Ricci scalar $R=R^i_{\phantom{i}i}$ \cite{KS}:
\begin{equation}
\mathfrak{L}=-\frac{1}{2\kappa}R\sqrt{-g}+\mathfrak{L}_\textrm{m},
\label{total}
\end{equation}
where $\mathfrak{L}$ is the total Lagrangian density, $R_{ik}=R^j_{\phantom{j}ijk}$ is the Ricci tensor, $R^i_{\phantom{i}mjk}=\Gamma^{\,\,i}_{m\,k,j}-\Gamma^{\,\,i}_{m\,j,k}+\Gamma^{\,\,i}_{l\,j}\Gamma^{\,\,l}_{m\,k}-\Gamma^{\,\,i}_{l\,k}\Gamma^{\,\,l}_{m\,j}$ is the curvature tensor, $\Gamma^{\,\,i}_{j\,k}$ is the affine connection, the comma denotes a partial derivative with respect to the coordinates, $g$ is the determinant of the metric tensor $g_{ik}$, $\mathfrak{L}_\textrm{m}$ is the Lagrangian density for matter, and $\kappa=8\pi G/c^4$ is Einstein's gravitational constant.
The metricity condition $g_{ij;k}=0$, where the semicolon denotes a covariant derivative with respect to the affine connection, gives the affine connection $\Gamma^{\,\,k}_{i\,j}=\{^{\,\,k}_{i\,j}\}+C^k_{\phantom{k}ij}$, where $\{^{\,\,k}_{i\,j}\}=(1/2)g^{km}(g_{mi,j}+g_{mj,i}-g_{ij,m})$ are the Christoffel symbols,
\begin{equation}
C^i_{\phantom{i}jk}=S^i_{\phantom{i}jk}+2S_{(jk)}^{\phantom{(jk)}i}
\label{contort}
\end{equation}
is the contortion tensor, $S^i_{\phantom{i}jk}=\Gamma^{\,\,\,\,i}_{[j\,k]}$ is the torsion tensor, $(\,)$ denotes symmetrization, and $[\,]$ denotes antisymmetrization.
The indices can be lowered with the metric tensor and raised with the contravariant metric tensor $g^{ik}$, as in general relativity \cite{LL}.

The curvature tensor can be decomposed as $R^i_{\phantom{i}klm}=P^i_{\phantom{i}klm}+C^i_{\phantom{i}km:l}-C^i_{\phantom{i}kl:m}+C^j_{\phantom{j}km}C^i_{\phantom{i}jl}-C^j_{\phantom{j}kl}C^i_{\phantom{i}jm}$, where $P^i_{\phantom{i}klm}$ is the Riemann tensor (the curvature tensor constructed from the Christoffel symbols instead of the affine connection) and the colon denotes a covariant derivative with respect to the Christoffel symbols.
We use the notation of \cite{Niko}.
The Ricci scalar can be decomposed as
\begin{equation}
R=P-4S^i_{\phantom{i}:i}-4S^i S_i-C^{ijk}C_{kij},
\label{decomp}
\end{equation}
where $P=P^j_{\phantom{j}ijk}g^{ik}$ is the Riemann scalar and
\begin{equation}
S_i=S^k_{\phantom{k}ik}
\end{equation}
is the torsion vector.

We consider the following Lagrangian density for matter:
\begin{equation}
\mathfrak{L}_\textrm{m}=\alpha S^i\phi_{,i}\sqrt{-g},
\label{Lagr}
\end{equation}
where $\phi$ is a scalar field and $\alpha$ is a real constant.
Varying the Lagrangian density (\ref{total}) with respect to the torsion tensor and equaling this variation to zero gives the Cartan field equations:
\begin{equation}
S^j_{\phantom{j}ik}-S_i \delta^j_k+S_k \delta^j_i=-\frac{\kappa}{2}s^{\phantom{ik}j}_{ik},
\end{equation}
where
\begin{equation}
s_i^{\phantom{i}jk}=\frac{2}{\sqrt{-g}}\frac{\delta\mathfrak{L}_\textrm{m}}{\delta C^i_{\phantom{i}jk}}
\label{spin}
\end{equation}
is the spin tensor of matter.
The inverse relation is
\begin{equation}
S^i_{\phantom{i}jk}=-\frac{\kappa}{2}(s_{jk}^{\phantom{jk}i}+\delta^i_{[j}s_{k]l}^{\phantom{k]l}l})
\label{Cartan}
\end{equation}
and its contraction gives the torsion vector:
\begin{equation}
S_i=\frac{1}{4}\kappa s_{ik}^{\phantom{ik}k}.
\end{equation}

For the Lagrangian density (\ref{Lagr}), the spin tensor defined by (\ref{spin}) is\footnote{
The definition of the spin tensor is rather formal in this case.
This quantity does not represent the spin of a scalar field.
}
\begin{equation}
s_{ij}^{\phantom{ij}k}=\frac{\alpha}{2}(\delta^k_i\phi_{,j}-\delta^k_j\phi_{,i}).
\end{equation}
Its contraction is thus
\begin{equation}
s_{ik}^{\phantom{ik}k}=-\frac{3\alpha}{2}\phi_{,i}.
\end{equation}
Substituting the spin tensor and its contraction to the field equations (\ref{Cartan}) gives the torsion tensor:
\begin{equation}
S^i_{\phantom{i}jk}=\frac{\alpha\kappa}{8}(\delta^i_j\phi_{,k}-\delta^i_k\phi_{,j}),
\end{equation}
and the torsion vector:
\begin{equation}
S_i=-\frac{3\alpha\kappa}{8}\phi_{,i}.
\end{equation}
The contortion tensor (\ref{contort}) is thus
\begin{equation}
C_{ijk}=\frac{\alpha\kappa}{4}(\phi_{,i}g_{jk}-\phi_{,j}g_{ik}).
\end{equation}
Substituting the torsion vector and contortion tensor to (\ref{total}), using (\ref{decomp}) and (\ref{Lagr}),
and omitting a covariant divergence which does not contribute to the field equations, gives the metric Lagrangian density:
\begin{equation}
\mathfrak{L}=-\frac{1}{2\kappa}P\sqrt{-g}-\frac{3\alpha^2\kappa}{32}\phi_{,i}\phi_{,k}g^{ik}\sqrt{-g}.
\label{Einstein}
\end{equation}
The second term on the right-hand side of this equation is the Lagrangian density for matter in the general-relativistic equivalent formulation of the Einstein--Cartan theory.
To obtain the Einstein field equations, the Lagrangian density (\ref{Einstein}) must be varied with respect to the metric tensor and such a variation must be equaled to zero.

The second term on the right-hand side of (\ref{Einstein}) has a form of a negative kinetic term in the Lagrangian density for the scalar field $\phi$, regardless of the sign of $\alpha$ in (\ref{Lagr}).
A similar result for a conformally coupled scalar field in the Einstein--Cartan gravity was described in \cite{conf}.
Such a phantom scalar field \cite{pha} could be a source of dark energy \cite{dark}.
It could also lead to the big rip of the universe \cite{rip}.
Furthermore, the Lagrangian density (\ref{Lagr}) is dynamically equivalent (differs by a covariant divergence) to a term in which the field $\phi$ is coupled to the covariant divergence of the torsion vector and has a nonkinetic form:
\begin{equation}
-\alpha S^i_{\phantom{i}:i}\phi\sqrt{-g}=-\alpha(S^i_{\phantom{i};i}-2S^i S_i)\phi\sqrt{-g}.
\end{equation}
Therefore, a kinetic scalar field can be generated from a nonkinetic scalar field by torsion.
This simple mathematical exercise suggests that dynamical scalar fields, whose equations of motion contain the second derivatives of the field with respect to time, may emerge from torsion.

\acknowledgments
This work was funded by the University Research Scholar program at the University of New Haven.

\end{document}